\documentclass[aps, showpacs, showkeys,nofootinbib,floatfix]{revtex4}

\usepackage{amssymb}
\usepackage{amsmath}
\usepackage{graphicx}
\usepackage[dvipdfm,colorlinks,citecolor=blue,urlcolor=blue]{hyperref}

\begin{document}
\title{\textbf{Energy Conditions and Stability in generalized $f(R)$ gravity with arbitrary coupling between matter and geometry}}

\author{Jun Wang$^1$ \footnote{E-mail: wangjun\_3@126.com}, Ya-Bo Wu$^1$
 \footnote{Corresponding author:ybwu61@163.com}, Yong-Xin Guo$^2$£¬Wei-Qiang Yang$^1$£¬Lei Wang$^1$}

\affiliation{$^1$Department of Physics, Liaoning Normal
University, Dalian 116029, P.R.China, $^2$College of Physics, Liaoning University, Shenyang 110036,
P.R.China}

\begin{abstract}
The energy conditions and the
Dolgov-Kawasaki criterion in generalized $f(R)$ gravity with
arbitrary coupling between matter and geometry are derived in this
paper, which are quite general and can degenerate to the well-known
energy conditions in GR and $f(R)$ gravity with non-minimal coupling
and non-coupling as special cases. In order to get some insight on
the meaning of these energy conditions and the Dolgov- Kawasaki
criterion, we apply them to a class of models in the FRW cosmology
and give some corresponding results.
\end{abstract}


\maketitle

\section{$\text{Introduction}$}
~~Recent astrophysical observations\cite{1,2} have indicated that
the expansion of our Universe is accelerating at the present time.
In principle, this phenomenon can be explained by either dark energy
(see, for instance, Ref.\cite{3} for reviews), in which the reason
of this phenomenon is due to an exotic component with large negative
pressure, or modified theories of gravity\cite{4}. Unfortunately, up
to now a satisfactory answer to the question that what dark energy
is and where it came from has not yet to be obtained. Alternative to
dark energy, modified theories of gravity is extremely attractive
because the cosmic speed-up can be easily explained by the fact that
some sub-dominant terms, like $1/R$, may become essential at small
curvature. Under some additional conditions, the early-time
inflation and late-time acceleration can be unified by different
role of gravitational terms relevant at small and at large
curvature.

$f(R)$ gravity is one of the competitive candidates in modified
theories of gravity (see, for instance, Refs.\cite{5,6} for
reviews). Here $f(R)$ is an arbitrary function of the Ricci scalar
$R$. One can add any form of $R$ in it, such as $1/R$\cite{7}, $\ln
R$\cite{8}, positive and negative powers of $R$\cite{9},
Gauss-Bonnet invariant\cite{10}, etc. It worth stressing that
considering some additional conditions, the early-time inflation and
late-time acceleration can be unified by different role of
gravitational terms relevant at small and at large curvature.
However, $f(R)$ gravity is not perfect because of containing a
number of instabilities. For instance, the theory with $1/R$ may
develop the instability\cite{22}. But by adding a term of $R^{2}$ to
this specific $f(R)$ model, one can remove this
instability\cite{8,9}. For more general forms of $f(R)$, the
stability condition $f'' \geq 0$ can be used to test $f(R)$ gravity
models\cite{25}.

This paper is organized as follows. In section 2, we give some
fundamental elements of generalized $f(R)$ gravity models with
arbitrary matter-geometry coupling. In section 3, the well-known
energy conditions, namely, the strong energy condition (SEC), the
null energy condition (NEC), the weak energy condition (WEC) and the
dominant energy condition (DEC), in the generalized $f(R)$ gravity
models, will be derived. In order to get some insight on the meaning
of these energy conditions, we apply them to a class of models.
Furthermore we rewritten them in terms of parameters of the
deceleration $(q)$, the jerk $(j)$ and the snap $(s)$ and then use
the rewritten WEC to restrict a special $f(R)$ model. The
instability of generalized $f(R)$ gravity models with arbitrary
matter-geometry coupling will be studied in section 4. Last section
contains our summary.

\section{Generalized $f(R)$ gravity models with arbitrary matter-geometry
coupling}
 A more general model of $f(R)$
gravity, in which the coupling style between matter and geometry is
arbitrary and the Lagrangian density of matter only appears in
coupling term, has been proposed in Ref.\cite{11}. Its starting
action is
\begin{equation}\label{1}
S=\int[\frac{1}{2}f_{1}(R)+G(L_{m})f_{2}(R)]\sqrt{-g}d^{4}x,
\end{equation}
where we have chosen $\kappa = 8\pi G= c = 1$. $f_{i}(R)$ ($i = 1,
2$) and $G(L_{m})$ are arbitrary functions of the Ricci scalar $R$
and the Lagrangian density of matter respectively. When $f_{2}(R) =
1$ and $G(L_{m}) = L_{m}$, we obtain the general form of $f(R)$
gravity with non-coupling between matter and geometry. Furthermore,
by setting $f_{1}(R) = R$, action (\ref{1}) can be reduced to the
standard General Relativity (GR).

Varying the action (\ref{1}) with respect to the metric $g_{\mu\nu}$
yields the field equations
\begin{equation}\label{2}
\begin{array}{rcl}
&
&F_{1}(R)R_{\mu\nu}-\frac{1}{2}f_{1}(R)g_{\mu\nu}+(g_{\mu\nu}\square-\triangledown_{\mu}\triangledown_{\nu})F_{1}(R)
=-2G(L_{m})F_{2}(R)R_{\mu\nu}\\ & &
-2(g_{\mu\nu}\square-\triangledown_{\mu}\triangledown_{\nu})
G(L_{m})F_{2}(R) -f_{2}(R)[K(L_{m})L_{m}-G(L_{m})]g_{\mu\nu}
\\ & & +f_{2}(R)K(L_{m})T_{\mu\nu},
\end{array}
\end{equation}
where $\Box=g^{\mu\nu}\nabla_{\mu}\nabla_{\nu}$, $F_{i}(R) =
df_{i}(R)/dR$ $(i = 1, 2)$ and $K (L_{m}) = dG(L_{m}) /dL_{m}$
respectively. The energy-momentum tensor of matter is defined as:
\begin{equation}\label{3}
T_{\mu\nu}=-\frac{2}{\sqrt{-g}}\frac{\delta(\sqrt{-g}L_{m})}{\delta
g^{\mu\nu}}.
\end{equation}
In this class of models, the energy-momentum tensor of matter is
generally not conserved due to the appearance of an extra
force\cite{12}.

\section{Energy conditions in the generalized $f(R)$ gravity models  with
arbitrary coupling between matter and geometry}
\subsection{The Raychaudhuri Equation}

Many models of $f(R)$ gravity have been proposed, which can be
restricted by imposing the so-called energy conditions\cite{13}.
These energy conditions were used in different contexts to derive
general results that hold for a variety of situations. Under these
energy conditions, one allows not only to establish gravity which
remains attractive, but also to keep the demands that the energy
density is positive and cannot flow faster than light. Below, we
simply review the Raychaudhuri equation which is the physical origin
of the NEC and the SEC\cite{14}.

In the case of a congruence of timelike geodesics defined by the
vector field $u^{\mu}$, the Raychaudhuri equation is given by
\begin{equation}\label{4}
\frac{d\theta}{d\tau}=-\frac{1}{3}\theta^{2}-\sigma_{\mu\nu}\sigma^{\mu\nu}+\omega_{\mu\nu}\omega^{\mu\nu}-R_{\mu\nu}u^{\mu}u^{\nu},
\end{equation}
where $R_{\mu\nu}$, $\theta$, $\sigma_{\mu\nu}$ and
$\omega_{\mu\nu}$ are the Ricci tensor, the expansion parameter, the
shear and the rotation associated with the congruence respectively.
While in the case of a congruence of null geodesics defined by the
vector field $k^{\mu}$, the Raychaudhuri equation is given by
\begin{equation}\label{5}
\frac{d\theta}{d\tau}=-\frac{1}{2}\theta^{2}-\sigma_{\mu\nu}\sigma^{\mu\nu}+\omega_{\mu\nu}\omega^{\mu\nu}-R_{\mu\nu}k^{\mu}k^{\nu}.
\end{equation}

From above expressions, it is clear that the Raychaudhuri equation
is purely geometric and independent of the gravity theory. In order
to constrain the energy-momentum tensor by the Raychaudhuri
equation, one can use the Ricci tensor from the field equations of
gravity to make a connection. Namely, through the combination of the
field equations of gravity and the Raychaudhuri equation, one can
obtain physical conditions for the energy-momentum tensor. Since
$\sigma^{2}\equiv\sigma_{\mu\nu}\sigma^{\mu\nu}\geq0$ (the shear is
a spatial tensor) and $\omega_{\mu\nu}=0$ (hypersurface orthogonal
congruence), from Eqs. (\ref{4}) and (\ref{5}), the conditions for
gravity to remain attractive ($d\theta/d\tau<0$) are
\begin{equation}\label{6}
R_{\mu\nu}u^{\mu}u^{\nu}\geq0  ~~~~~~~~~~SEC,
\end{equation}
\begin{equation}\label{7}
R_{\mu\nu}k^{\mu}k^{\nu}\geq0  ~~~~~~~~~~NEC.
\end{equation}

Thus by means of the relationship (\ref{6}) and Einstein's equation,
one obtains
\begin{equation}\label{8}
R_{\mu\nu}u^{\mu}u^{\nu}=(T_{\mu\nu}-\frac{T}{2}g_{\mu\nu})u^{\mu}u^{\nu}\geq0,
\end{equation}
where $T_{\mu\nu}$ is the energy-momentum tensor and $T$ is its
trace. If one considers a perfect fluid with energy density $\rho$
and pressure $p$,
\begin{equation}\label{9}
T_{\mu\nu}=(\rho+p)U_{\mu}U_{\nu}-pg_{\mu\nu},
\end{equation}
the relationship (\ref{8}) turns into the well-known SEC of
Einstein's theory, i.e.,
\begin{equation}\label{10}
\rho+3p\geq0.
\end{equation}

Similarly, by using the relationship (\ref{7}) and Einstein's
equation, one has
\begin{equation}\label{11}
T_{\mu\nu}k^{\mu}k^{\nu}\geq0.
\end{equation}
Then considering Eq.(\ref{9}), the familiar NEC of general
relativity can be reproduced as:
\begin{equation}\label{12}
\rho+p\geq0.
\end{equation}

\subsection{Energy conditions}
~~The Einstein
tensor resulting from the field equations (\ref{2}) is
\begin{equation}\label{13}
G_{\mu\nu}\equiv R_{\mu\nu}-\frac{1}{2}g_{\mu\nu}R=T_{\mu\nu}^{eff},
\end{equation}
where the effective energy-momentum tensor $T_{\mu\nu}^{eff}$ is
defined as follows:
\begin{equation}\label{14}
\begin{array}{rcl}
T_{\mu\nu}^{eff} & = &
\frac{1}{f'_{1}+2Gf'_{2}}\{\frac{1}{2}g_{\mu\nu}[f_{1}-(f'_{1}+2Gf'_{2})R]-
(g_{\mu\nu}\Box-\nabla_{\mu}\nabla_{\nu})f'_{1}\\
& &
-2(g_{\mu\nu}\Box-\nabla_{\mu}\nabla_{\nu})Gf'_{2}-f_{2}(G'L_{m}-G)g_{\mu\nu}+f_{2}G'T_{\mu\nu}\},
\end{array}
\end{equation}
where $f_{i}=f_{i}(R)$ ($i=1,2$), $G=G(L_{m})$ and the prime denotes
differentiation with respect to the Ricci scalar $R$ and the
Lagrangian density $L_{m}$ respectively. Contracting the above
equation, we have
\begin{equation}\label{15}
\begin{array}{rcl}
T^{eff} & = &
\frac{1}{f'_{1}+2Gf'_{2}}\{2[f_{1}-(f'_{1}+2Gf'_{2})R]-3\Box
f'_{1}\\ & & -6\Box Gf'_{2}-4f_{2}(G'L_{m}-G)+f_{2}G'T\},
\end{array}
\end{equation}
where $T=g^{\mu\nu}T_{\mu\nu}$. Thus, we can write $R_{\mu\nu}$ in
terms of an effective stress-energy tensor and its trace, i.e.,
\begin{equation}\label{16}
R_{\mu\nu}=T_{\mu\nu}^{eff}-\frac{1}{2}g_{\mu\nu}T^{eff}.
\end{equation}

In order to keep gravity attractive, besides the expressions
(\ref{6}) and (\ref{7}), the following additional condition should
be required
\begin{equation}\label{17}
\frac{f_{2}G'}{f'_{1}+2Gf'_{2}}>0.
\end{equation}
Note that this condition is independent of the ones derived from the
Raychaudhuri equation (i.e., the expressions (\ref{6}) and
(\ref{7})), and only relates to an effective gravitational coupling.

The FRW metric is chosen as:
\begin{equation}\label{18}
ds^{2}=dt^{2}-a^{2}(t)ds^{2}_{3},
\end{equation}
where $a(t)$ is the scale factor and $ds^{2}_{3}$ contains the
spacial part of the metric. Using this metric, we can obtain
$R=-6(2H^{2}+\dot{H})$, where $H=\dot{a}(t)/a(t)$ is the Hubble
expansion parameter, and
$\Gamma^{0}_{\mu\nu}=a(t)\dot{a}(t)\delta_{\mu\nu} ~(\mu, \nu\neq0)$
are the components of the affine connection.

By using the relationship (\ref{6}) and Eq. (\ref{16}), the SEC can
be given as:
\begin{equation}\label{19}
T_{\mu\nu}^{eff}u^{\mu}u^{\nu}-\frac{1}{2}T^{eff}\geq0,
\end{equation}
where we have used the condition $g_{\mu\nu}u^{\mu}u^{\nu}=1$.
Taking the energy-momentum tensor $T_{\mu\nu}$ to be a perfect fluid
(i.e., Eq.(\ref{9})) and considering the condition (\ref{17}), we
obtain
\begin{equation}\label{20}
\begin{array}{rcl}& &
\rho+3p-\frac{1}{f_{2}G'}[f_{1}-(f'_{1}+2Gf'_{2})R]+3\frac{f''_{1}}{f_{2}G'}(H\dot{R}+\ddot{R})\\
& & +3\frac{f'''_{1}}{f_{2}G'}\dot{R}^{2}
+6\frac{1}{f_{2}G'}(G''\dot{L_{m}}^{2}f'_{2}+\ddot{L_{m}}G'f'_{2}+2f''_{2}\dot{R}G'\dot{L_{m}}\\
& & +f'''_{2}\dot{R}^{2}G+f''_{2}\ddot{R}G)
+6\frac{H}{f_{2}G'}(G'\dot{L_{m}}f'_{2}+f''_{2}\dot{R}G)
\\ & & +\frac{2}{G'}(G'L_{m}-G)\geq0,
\end{array}
\end{equation}
where the dot denotes differentiation with respect to cosmic time.
This is the SEC in $f(R)$ gravity with arbitrary coupling between
matter and geometry.

The NEC in $f(R)$ gravity with arbitrary coupling between matter and
geometry can be expressed as:
\begin{equation}\label{21}
T_{\mu\nu}^{eff}k^{\mu}k^{\nu}\geq0.
\end{equation}
By the same method as the SEC, the above relationship can be changed
into
\begin{equation}\label{22}
\begin{array}{rcl}
& &
\rho+p+(H\dot{R}+\ddot{R})\frac{f''_{1}}{f_{2}G'}+\frac{f'''_{1}}{f_{2}G'}\dot{R}^{2}+\frac{2}{f_{2}G'}
(G''\dot{L_{m}}^{2}f'_{2}+\\ & & \ddot{L_{m}}G'f'_{2}+
2f''_{2}\dot{R}G'\dot{L_{m}}+f'''_{2}\dot{R}^{2}G+f''_{2}\ddot{R}G)-\\
& & \frac{2H}{f_{2}G'}(G'\dot{L_{m}}f'_{2}+f''_{2}\dot{R}G)\geq0.
\end{array}
\end{equation}

From above discussions, it is worth stressing that by taking
$G(L_{m})=L_{m}$ and rescaling the function $f_{2}(R)$ as $1+\lambda
f_{2}(R)$ in expressions (\ref{20}) and (\ref{22}), we can obtain
the SEC and the NEC in $f(R)$ gravity with non-minimal coupling
between matter and geometry, which are just the results given in
Ref.\cite{15}. While by setting $f_{2}(R)=1$ and $G(L_{m})=L_{m}$,
we can derive the SEC and the NEC in $f(R)$ gravity with
non-coupling, which are just the same as the ones in Ref.\cite{16}.
Furthermore, when $f_{1}(R)=R$, the SEC and the NEC in general
relativity, i.e., $\rho+3p\geq0$ and $\rho+p\geq0$, can be
reproduced.

Note that the above expressions of the SEC and the NEC are directly
derived from Raychaudhuri equation. However, equivalent results can
be obtained by taking the transformations $ \rho \rightarrow
\rho^{eff}$ and $p\rightarrow p^{eff}$ into $\rho+3p\geq0$ and
$\rho+p\geq0$. Thus by extending this approach to $\rho-p\geq0$ and
$\rho\geq0$, we will give the DEC and the WEC in $f(R)$ gravity with
arbitrary coupling between matter and geometry in the following.

By means of Eqs. (\ref{14}) and (\ref{18}), the effective energy
density and the effective pressure can be derived as follows:
\begin{equation}\label{23}
\begin{array}{rcl}
\rho^{eff} & = &
\frac{1}{f'_{1}+2Gf'_{2}}\{\frac{1}{2}[f_{1}-(f'_{1}+2Gf'_{2})R]-3H\dot{R}f''_{1}-6H(G'\dot{L_{m}}f'_{2}+
\\ & & f''_{2}\dot{R}G) -f_{2}(G'L_{m}-G)+f_{2}G'\rho\},
\end{array}
\end{equation}

\begin{equation}\label{24}
\begin{array}{rcl}
p^{eff}& = &
\frac{1}{f'_{1}+2Gf'_{2}}\{-\frac{1}{2}[f_{1}-(f'_{1}+2Gf'_{2})R]+(2H\dot{R}+\ddot{R})f''_{1}+f'''_{1}\dot{R}^{2}
\\ & &
+2(G''\dot{L_{m}}^{2}f'_{2}+\ddot{L_{m}}G'f'_{2}+2f''_{2}\dot{R}G'\dot{L_{m}}+f'''_{2}\dot{R}^{2}G+f''_{2}\ddot{R}G)
\\ & &
4H(G'\dot{L_{m}}f'_{2}+f''_{2}\dot{R}G)+f_{2}(G'L_{m}-G)+f_{2}G'p\}.
\end{array}
\end{equation}
Then, the corresponding DEC and WEC in $f(R)$ gravity with arbitrary
coupling can be respectively written as:
\begin{equation}\label{25}
\begin{array}{rcl}& &
\rho-p+\frac{1}{f_{2}G'}[f_{1}-(f'_{1}+2Gf'_{2})R]-(5H\dot{R}+\ddot{R})\frac{f''_{1}}{f_{2}G'}-\\
& & \frac{f'''_{1}}{f_{2}G'}\dot{R^{2}} -\frac{2}{f_{2}G'}
(G''\dot{L_{m}}^{2}f'_{2}+\ddot{L_{m}}G'f'_{2}+2f''_{2}\dot{R}G'\dot{L_{m}}+
\\ & &
f'''_{2}\dot{R}^{2}G+f''_{2}\ddot{R}G)-\frac{10H}{f_{2}G'}
(G'\dot{L_{m}}f'_{2}+f''_{2}\dot{R}G)-\frac{2}{G'}\\ & &
(G'L_{m}-G)\geq0,
\end{array}
\end{equation}

\begin{equation}\label{26}
\begin{array}{rcl}& &
\rho+\frac{1}{2f_{2}G'}[f_{1}-(f'_{1}+2Gf'_{2})R]-3H\dot{R}\frac{f''_{1}}{f_{2}G'}-6H\frac{1}{f_{2}G'}
\\ & &
(G'\dot{L_{m}}f'_{2}+f''_{2}\dot{R}G)-\frac{1}{G'}(G'L_{m}-G)\geq0.
\end{array}
\end{equation}

We show that by taking $G(L_{m})=L_{m}$ and rescaling the function
$f_{2}(R)$ as $1+\lambda f_{2}(R)$, above expressions are the DEC
and the WEC in $f(R)$ gravity with non-minimal coupling between
matter and geometry, which are just the same as the ones in
Ref.\cite{15}. While by setting $f_{2}(R)=1$ and $G(L_{m})=L_{m}$,
the results given by us are the DEC and the WEC in $f(R)$ gravity
with non-coupling, which are consistent with the results given in
Ref.\cite{16}. Furthermore, when $f_{1}(R)=R$, the DEC  and the WEC
in general relativity, i.e., $\rho-p\geq0$ and $\rho\geq0$, can be
reproduced.

\subsection{Energy Conditions for a Class of Models}
~~In order to get some insight on the meaning of the above energy
conditions, we consider a specific type of models where $f_{1}(R)$
and $f_{2}(R)$ are taken as
\begin{equation}\label{m1}
\begin{array}{rcl}& &
f_{1}(R)=R+\epsilon R^{n},\\& & f_{2}(R)=\alpha R^{m},
\end{array}
\end{equation}
In the FRW cosmology, the energy conditions can be written as
\begin{equation}\label{m2}
\frac{\hat{\epsilon}\mid R\mid^{n}}{\hat{\alpha} \mid
R\mid^{m}G'(L_{m})}\{\frac{2\hat{\alpha}}{\hat{\epsilon}}[G(L_{m})C_{m}+A]\mid
R\mid^{m-n}+C_{n}\}\geq B,
\end{equation}
where $A$,$B$ and $C_{m,n}$ depend on the energy condition under
study and we take $\hat{\epsilon}=(-1)^{n}\epsilon$ and
$\hat{\alpha}=(-1)^{n}\alpha$ due to the fact that for a FRW metric
one has $R<0$. For the SEC, one finds
\begin{subequations}
\begin{equation}\label{m3a}
\begin{array}{rcl}
A^{SEC} & = &
G'(L_{m})[L_{m}+3mR^{-1}(\ddot{L_{m}}+H\dot{L_{m}})+6m\dot{L_{m}}\dot{R}R^{-2}(m-1)]\\&
&+3m\dot{L_{m}}^{2}G''(L_{m})R^{-1},
\end{array}
\end{equation}

\begin{equation}\label{m3b}
B^{SEC}=-(\rho+3p),
\end{equation}

\begin{equation}\label{m3c}
\begin{array}{rcl}
C^{SEC}_{n} & = &
(n-1)[3\ddot{R}nR^{-2}+1+3HnR^{-2}\dot{R}+3nR^{-3}\dot{R}^{2}(n-2)].
\end{array}
\end{equation}
\end{subequations}

For the NEC, one obtains
\begin{subequations}
\begin{equation}\label{m4a}
\begin{array}{rcl}
A^{NEC} & = &
[m\ddot{L_{m}}G'(L_{m})R^{-1}-Hm\dot{L_{m}}G'(L_{m})R^{-1}\\&
&+2m\dot{L_{m}}\dot{R}G'(L_{m})R^{-2}(m-1)]+m\dot{L_{m}}G''(L_{m})R^{-1},
\end{array}
\end{equation}

\begin{equation}\label{m4b}
\begin{array}{rcl}
B^{NEC}=-(\rho+p),
\end{array}
\end{equation}

\begin{equation}\label{m4c}
\begin{array}{rcl}
C^{NEC}_{n} & = &
(n-1)n[\ddot{R}R^{-2}+HR^{-2}\dot{R}+R^{-3}\dot{R}^{2}(n-2)].
\end{array}
\end{equation}
\end{subequations}

For the DEC, one has
\begin{subequations}
\begin{equation}\label{m5a}
\begin{array}{rcl}
A^{DEC} & = &
G'(L_{m})[-L_{m}-m\ddot{L_{m}}R^{-1}-5Hm\dot{L_{m}}R^{-1}\\&
&+(1-m)2m\dot{L_{m}}\dot{R}R^{-2}]-m\dot{L_{m}}^{2}G''(L_{m})R^{-1},
\end{array}
\end{equation}

\begin{equation}\label{m5b}
\begin{array}{rcl}
B^{DEC}=-(\rho-p),
\end{array}
\end{equation}

\begin{equation}\label{m5c}
\begin{array}{rcl}
C^{DEC}_{n} & = &
(1-n)[\ddot{R}nR^{-2}+1+5HnR^{-2}\dot{R}+nR^{-3}\dot{R}^{2}(n-2)].
\end{array}
\end{equation}
\end{subequations}

Finally, for the WEC, one gets
\begin{subequations}
\begin{equation}\label{m6a}
\begin{array}{rcl}
A^{WEC} & = & -L_{m}-6Hm\dot{L_{m}}R^{-1},
\end{array}
\end{equation}

\begin{equation}\label{m6b}
\begin{array}{rcl}
B^{WEC}=-\rho,
\end{array}
\end{equation}

\begin{equation}\label{m6c}
\begin{array}{rcl}
C^{WEC}_{n} & = & (1-n)(\frac{1}{2}+3HnR^{-2}\dot{R}).
\end{array}
\end{equation}
\end{subequations}

Given these definitions, the study of all the energy conditions can
be performed by satisfying the inequality (\ref{m2}). Note that all
the energy conditions depend on the geometrical parameters. It means
that for different models, the energy conditions can be satisfied by
choosing them properly.

For models given by Eq.(\ref{m1}), the condition for keeping gravity
attractive(GA), i.e. inequality (\ref{17}), also can be obtained
from inequality (\ref{m2}) by taking
\begin{subequations}
\begin{equation}\label{m7a}
\begin{array}{rcl}
A^{GA} & = & \frac{1}{2\hat{\alpha}}\mid R\mid^{-m},
\end{array}
\end{equation}

\begin{equation}\label{m7b}
\begin{array}{rcl}
B^{GA}=0,
\end{array}
\end{equation}

\begin{equation}\label{m7c}
\begin{array}{rcl}
C^{GA}_{n} & = & nR^{-1}.
\end{array}
\end{equation}
\end{subequations}
This means that inequality (\ref{m2}) also stands for the condition
that ensures gravity remains attractive for models given by
Eq.(\ref{m1}).

In the following, we use energy conditions to restrict a special
$f(R)$ model also in the FRW cosmology. The Ricci scalar $R$ and its
derivatives can be expressed by the parameters of the deceleration
$(q)$, the jerk $(j)$ and the snap $(s)$\cite{17}, namely,
\begin{subequations}
\begin{equation}\label{27a}
R=-6H^{2}(1-q),
\end{equation}
\begin{equation}\label{27b}
\dot{R}=-6H^{3}(j-q-2),
\end{equation}
\begin{equation}\label{27c}
\ddot{R}=-6H^{4}(s+q^{2}+8q+6),
\end{equation}
\end{subequations}
where
\begin{equation}\label{28}
q=-\frac{1}{H^{2}}\frac{\ddot{a}}{a},
~~j=\frac{1}{H^{3}}\frac{\stackrel{...}{a}}{a}, ~~and
~~s=\frac{1}{H^{4}}\frac{\stackrel{....}{a}}{a}.
\end{equation}
Thus, the energy conditions (\ref{20}), (\ref{22}), (\ref{25}) and
(\ref{26}) can be rewritten as:

\begin{subequations}
\begin{equation}\label{29a}
\begin{array}{rcl}
& &
\rho+3p-\frac{1}{f_{2}G'}[f_{1}+6H^{2}(f'_{1}+2Gf'_{2})(1-q)]-18\frac{f''_{1}}{f_{2}G'}H^{4}(j+s+q^{2}+7q+4)\\
& & +108\frac{f'''_{1}}{f_{2}G'}H^{6}(j-q-2)^{2}
+6\frac{1}{f_{2}G'}[G''\dot{L_{m}}^{2}f'_{2}+\ddot{L_{m}}G'f'_{2}-12f''_{2}H^{3}(j-q\\
& &
-2)G'\dot{L_{m}}+36f'''_{2}H^{6}(j-q-2)^{2}G-6f''_{2}H^{4}(s+q^{2}+8q+6)G]
+6\frac{H}{f_{2}G'}
\\ & &
[G'\dot{L_{m}}f'_{2}-6f''_{2}H^{3}(j-q-2)G]+\frac{2}{G'}(G'L_{m}-G)\geq0,~~~~~~~~~~(SEC)
\end{array}
\end{equation}

\begin{equation}\label{29b}
\begin{array}{rcl}
& &
\rho+p-6H^{4}(j+s+q^{2}+7q+4)\frac{f''_{1}}{f_{2}G'}+36\frac{f'''_{1}}{f_{2}G'}H^{6}(j-q-2)^{2}+\frac{2}{f_{2}G'}
\\ & & [G''\dot{L_{m}}^{2}f'_{2}+\ddot{L_{m}}G'f'_{2}
-12f''_{2}H^{3}(j-q-2)G'\dot{L_{m}}+36f'''_{2}H^{6}(j-q\\
& &
-2)^{2}G-6f''_{2}H^{4}(s+q^{2}+8q+6)G]-\frac{2H}{f_{2}G'}[G'\dot{L_{m}}f'_{2}-6f''_{2}H^{3}(j-q\\
& & -2)G]\geq0,~~~~~~(NEC)
\end{array}
\end{equation}

\begin{equation}\label{29c}
\begin{array}{rcl}
& &
\rho-p+\frac{1}{f_{2}G'}[f_{1}+6H^{2}(f'_{1}+2Gf'_{2})(1-q)]+6H^{4}(5j+s+q^{2}+3q\\
& &
-4)\frac{f''_{1}}{f_{2}G'}-36\frac{f'''_{1}}{f_{2}G'}H^{6}(j-q-2)^{2}-\frac{2}{f_{2}G'}
[G''\dot{L_{m}}^{2}f'_{2}+\ddot{L_{m}}G'f'_{2}-12f''_{2}H^{3}
\\ & &
(j-q-2)G'\dot{L_{m}}+36f'''_{2}H^{6}(j-q-2)^{2}G-6f''_{2}H^{4}(s+q^{2}+8q+6)G]
\\ & &
-\frac{10H}{f_{2}G'}
[G'\dot{L_{m}}f'_{2}-6f''_{2}H^{3}(j-q-2)G]-\frac{2}{G'}(G'L_{m}-G)\geq0,~~~~(DEC)
\end{array}
\end{equation}

\begin{equation}\label{29d}
\begin{array}{rcl}
& &
\rho+\frac{1}{2f_{2}G'}[f_{1}+6H^{2}(f'_{1}+2Gf'_{2})(1-q)]+18H^{4}(j-q-2)\frac{f''_{1}}{f_{2}G'}-6H\frac{1}{f_{2}G'}
\\ & &
[G'\dot{L_{m}}f'_{2}-6f''_{2}H^{3}(j-q-2)G]-\frac{1}{G'}(G'L_{m}-G)\geq0.~~~~(WEC)
\end{array}
\end{equation}
\end{subequations}

To exemplify how to use these energy conditions to constrain the
$f(R)$ theories of gravity, we consider a special model with
$f_{1}(R)=R$, $f_{2}(R)=\alpha R^{n}$ and
$G(L_{m})=L_{m}=-\rho$\cite{18}. Since there has been no reliable
measurement for the snap parameter $(s)$ up to now, we only focus on
the WEC. Under the requirement $f'(R)>0 $ for all $R$ and taking
$H_{0}=70.5$\cite{19}, the WEC (\ref{29d}) in this particular case
is
\begin{equation}\label{30}
\begin{array}{rcl}
0.3Bn^{2}-0.3n(1+B)+1\geq0,
\end{array}
\end{equation}
where $B=(j-q-2)/(1-q)^{2}$.

From the above expression, it is easy to see that the coefficient
$\alpha$ is arbitrary and the value of the index $n$ depends on $B$.
Taking $q_{0}=-0.81\pm0.14$ and
$j_{0}=2.16^{+0.81}_{-0.75}$\cite{20} (the subscript $0$ denotes the
present value), we can give the present range of $B$ is $0.03\leq B
_{0}\leq 0.5$. By calculations and analysis, the results of the
expression (\ref{30}) are as follows: when the real solution exists,
the range of $B$ is either $B \leq \frac{17-2\sqrt{70}}{3}$ or
$B\geq \frac{17+2\sqrt{70}}{3}$. Considering $0.03\leq B _{0}\leq
0.5$, we find the range of $B$ is $0.03\leq B \leq
\frac{17-2\sqrt{70}}{3}$ and the index $n$ are $7.36577 \leq
n_{+}\leq 30.716$ and $3.61737\leq n_{-}\leq5.26214$. However, when
there is not any real solution, the range of $B$ is
$\frac{17-2\sqrt{70}}{3}< B \leq 0.5$ and the index $n$ can be taken
as any real number.

We can point out that for the model of $f_{1}(R)=R+\alpha R^{n}$,
$f_{2}=1$ and $G(L_{m})=L_{m}$, the corresponding results to the WEC
given by us are just the same as the ones in Ref.\cite{16}.

\section{The instability of generalized $f(R)$ gravity models with arbitrary matter-geometry coupling}
~~Modified gravity must be stable at the classical and quantum
level. There are in principle several kinds of instabilities to
consider\cite{21}. Dolgov-Kawasaki instability\cite{22} is one of
them. Below, we will focus on this instability and generalize to
$f(R)$ gravity models with arbitrary matter-geometry coupling.

The trace of the field equation (\ref{2}) is

\begin{equation}\label{dk1}
\begin{array}{rcl}& &
R+\frac{1}{f'_{1}+2Gf'_{2}}\{2[f_{1}-(f'_{1}+2Gf'_{2})R]-3\Box
f'_{1}-6\Box Gf'_{2}-4f_{2}(G'L_{m}-G)\}\\&
&=\frac{-1}{f'_{1}+2Gf'_{2}}f_{2}G'T,
\end{array}
\end{equation}
where $T=g^{\mu\nu}T_{\mu\nu}$. As usual, we take $f_{1}(R)$ as
$f_{1}(R)=R+\epsilon\varphi(R)$, where $\epsilon$ must be small to
compatibility with Solar System experiment\cite{23}.
Following\cite{22}, we expand the space-time quantities of interest
as the sum of a background with constant curvature and a small
perturbation: $R=R_{0}+R_{1}$, $T=T_{0}+T_{1}$, $L=L_{0}+L_{1}$, and
the space-time metric can locally be approximated by
$g_{\mu\nu}=\eta_{\mu\nu}+h_{\mu\nu}$, where $\eta_{\mu\nu}$ is the
Minkowski metric. In fact, this is a local expansion over small
space-time regions that are locally flat. Accordingly,
$f_{1}(R)=R_{0}+R_{1}+\epsilon\varphi(R_{0})+\epsilon\varphi'(R_{0})R_{1}+...$,
$f'_{1}(R)=1+\epsilon\varphi'(R_{0})+\epsilon\varphi''(R_{0})R_{1}+...$
and the linearized version of the trace equation (\ref{dk1}) in the
perturbations yields
\begin{equation}\label{dk2}
\begin{array}{rcl}& &
[6G(L_{m})f_{2}''(R_{0})+3\epsilon\varphi''(R_{0})]\ddot{R_{1}}-3\epsilon\varphi''(R_{0})\nabla^{2}R_{1}+[12\dot{L_{0}}G'(L_{m})
f''_{2}(R_{0})\\& &+12\dot{L_{1}}G'(L_{m})
f''_{2}(R_{0})+12G(L_{m})\dot{R_{0}}f_{2}'''(R_{0})+6\epsilon\dot{R_{0}}\varphi'''(R_{0})]\dot{R_{1}}\\&
&-6\epsilon\varphi'''(R_{0})
\vec{\nabla}R_{1}\cdot\vec{\nabla}R_{0}-[1-2G(L_{m})f_{2}'(R_{0})+\epsilon\varphi'(R_{0})-R_{0}\epsilon\varphi''(R_{0})
\\&
&-3\ddot{R_{0}}\epsilon\varphi'''(R_{0})+3\epsilon\varphi'''(R_{0})\nabla^{2}R_{0}]R_{1}=6f_{2}'(R_{0})\nabla^{2}G(L_{m})
\\&
&-f_{2}(R_{0})(4L_{1}-T_{1})G'(L_{m})-6(\ddot{L_{0}}+\ddot{L_{1}})f_{2}'(R_{0})G'(L_{m})-[6G(L_{m})f_{2}''(R_{0})
\\&
&+3\epsilon\varphi''(R_{0})]\ddot{R_{0}}-12(\dot{L_{0}}+\dot{L_{1}})\dot{R_{0}}G'(L_{m})f_{2}''(R_{0})-12\dot{L_{0}}
\dot{L_{1}}f_{2}'(R_{0})G''(L_{m})\\&
&+3\epsilon\varphi''(R_{0})\nabla^{2}R_{0},
\end{array}
\end{equation}
where $\vec{\nabla}$, $\nabla^{2}$ and overdot denote the gradient,
Laplacian operators in Euclidean three-dimensional space and
differentiation with respect to time, respectively, and the zero
order equation
\begin{equation}\label{dk3}
\begin{array}{rcl}& &
f_{2}(R_{0})T_{0}G'(L_{m})=-4f_{2}(R_{0})G(L_{m})-2\epsilon\varphi(R_{0})+4G(L_{m})R_{0}f_{2}'(R_{0})\\&
&+4f_{2}(R_{0})L_{0}
G'(L_{m})+2R_{0}\epsilon\varphi'(R_{0})-R_{0}-2G(L_{m})R_{0}f_{2}'(R_{0})-R_{0}\epsilon\varphi'(R_{0})
\end{array}
\end{equation}
has been used. By further calculation, the effective mass $m_{eff}$
of the dynamical degree of freedom $R_{1}$ can be given as
\begin{equation}\label{dk4}
\begin{array}{rcl}
m_{eff}^{2}& =
&[6G(L_{m})f_{2}''(R_{0})+3\epsilon\varphi''(R_{0})]^{-1}
[2G(L_{m})f_{2}'(R_{0})-1-\epsilon\varphi'(R_{0})
\\&
&+R_{0}\epsilon\varphi''(R_{0})+3\ddot{R_{0}}\epsilon\varphi'''(R_{0})
-3\epsilon\varphi'''(R_{0})\nabla^{2}R_{0}].
\end{array}
\end{equation}
The dominant term on the right hand side is
$[6G(L_{m})f_{2}''(R_{0})+3\epsilon\varphi''(R_{0})]^{-1}$ and the
effective mass squared must be non-negative for stability.
Therefore, $f_{1}''(R)+2G(L_{m})f_{2}''(R)\geq0$ is the stability
criterion for the generalized $f(R)$ gravity models with arbitrary
matter-geometry coupling against Dolgov-Kawasaki instabilities.

Note that by taking $G(L_{m})=L_{m}$ and rescaling the function
$f_{2}(R)$ as $1+\lambda f_{2}(R)$, this criterion is the
Dolgov-Kawasaki criterion in $f(R)$ gravity with non-minimal
coupling between matter and geometry, which is just the same as the
one in Ref.\cite{24}. While by setting $f_{2}(R)=1$ and
$G(L_{m})=L_{m}$, the results given by us is the Dolgov-Kawasaki
criterion in $f(R)$ gravity with non-coupling, which is consistent
with the results given in Ref.\cite{25}.

For models given by Eq.(\ref{m1}), the Dolgov-Kawasaki criterion is
\begin{equation}\label{dk5}
\begin{array}{rcl}
\hat{\epsilon}n(n-1)R^{n}+2G(L_{m})\hat{\alpha} m(m-1)R^{m}\geq0
\end{array}
\end{equation}
where

\begin{equation}\label{dk6}
\begin{array}{rcl}& &
\hat{\epsilon}= \left\{%
\begin{array}{ll}
(-1)^{n}\epsilon, &\textrm{if }R<0\\
 ~~\epsilon, &\textrm{if }R>0
\end{array}%
\right.
,~~~~~~\hat{\alpha}= \left\{%
\begin{array}{ll}
(-1)^{n}\alpha, &\textrm{if }R<0\\
 ~~~\alpha, &\textrm{if }R>0
\end{array}%
\right. .\end{array}
\end{equation}
It is clear that the stability criterion of these models don't
relate to the values of $j$ and $s$, and the space-time only depends
on $R$. When $n=m$ the inequality (\ref{dk5}) gives,
$\epsilon+2G(L_{m})\alpha\geq0$.

It is worth stressing that the inequality (\ref{dk5}) also can be
obtained from the inequality (\ref{m2}) by taking
\begin{subequations}
\begin{equation}\label{dk7a}
\begin{array}{rcl}
A^{DK} & = & 0,
\end{array}
\end{equation}

\begin{equation}\label{dk7b}
\begin{array}{rcl}
B^{DK}& = & 0,
\end{array}
\end{equation}

\begin{equation}\label{dk7c}
\begin{array}{rcl}
C^{DK}_{n} & = & n(n-1)\hat{\alpha} \mid R\mid^{m}G'(L_{m}).
\end{array}
\end{equation}
\end{subequations}

From the above discussions, we find that for models given by
Eq.(\ref{m1}), the energy conditions, the Dolgov-Kawasaki criterion
and the condition for attractive gravity have the the same type of
inequalities, but note that they are independent each other.

From the inequality (\ref{dk5}), we see that the viability of the
model with respect to the Dolgov-Kawasaki instability criterion will
depend not only on the value of the constants $\epsilon$ and
$\alpha$, but also on the space-time metric under consideration.
This fact will give further constraints on the Ricci scalar.

\section{Summary}
~~So far, we have derived the energy conditions (SEC, NEC, DEC, WEC)
in the generalized $f(R)$ gravity models with arbitrary coupling
between matter and geometry. For the SEC and the NEC, the
Raychaudhuri equation, which is the physical origin of them, has
been used. From the derivation, we found that equivalent results can
be obtained by taking the transformations $ \rho \rightarrow
\rho^{eff}$ and $p\rightarrow p^{eff}$ into $\rho+3p\geq0$ and
$\rho+p\geq0$. Thus by extending this approach to $\rho-p\geq0$ and
$\rho\geq0$, the DEC and the WEC in the generalized $f(R)$ gravity
models with arbitrary coupling between matter and geometry can be
obtained. The condition to keep gravity attractive and the
Dolgov-Kawasaki criterion in the generalized $f(R)$ gravity models
have been also given, but the approaches of deriving them are
different.

It is worth noting that the energy conditions and the
Dolgov-Kawasaki criterion obtained in this paper are quite general,
which include the corresponding results given in
Refs.\cite{15,16,24,25} as well as in the general relativity (GR) as
special cases.

Furthermore, in order to get some insight on the meaning of these
energy conditions and the Dolgov-Kawasaki criterion, we have applied
them to a class of models. In these models the energy conditions,
the Dolgov-Kawasaki criterion and the condition for attractive
gravity have the the same type of inequalities. By analysis, we find
that the Dolgov-Kawasaki instability criterion depends not only on
the value of the constants $\epsilon$ and $\alpha$ but also on the
space-time metric under consideration.

In addition, we have considered the special model with $f_{1}(R)=R$,
$f_{2}(R)=\alpha R^{n}$ and $G(L_{m})=L_{m}=-\rho$. By virtue of the
WEC and the present astrophysical observations, the values of
parameters $\alpha$ and $n$ can be constrained in this model. Of
course, we will continue to study other models of $f(R)$ gravity
with arbitrary coupling between matter and geometry in our following
investigations.

\section*{Acknowledgements}
~~This work was partly supported by the National Natural Science
Foundation of China (Grant Nos.$10875056$,$10932002$ and
$10872084$), and the Scientific Research Foundation of the Higher
Education Institute of Liaoning Province, China (Grant
Nos.$2007T087$ and $2009R35$).


\begin{thebibliography}{10} \small
\bibitem[1]{1}A.G. Riess, et al., Astron. J. 116 (1998) 1009.
\bibitem[2]{2}S. Perlmutter, et al., Astrophys. J. 517 (1999) 565.
\bibitem[3]{3}E.J. Copeland, M. Sami, S. Tsujikawa, \href{http://arxiv.org/abs/hep-th/0603057}{arXiv: 0603057} [hep-th].
\bibitem[4]{4}C. Brans, R.H. Dicke, Phys. Rev. 124 (1961) 925;\\
V. Faraoni, Cosmology in Scalar-Tensor Gravity (Kluwer Academic,
Dordrecht) 2004;\\Dvali, et al., Phys. Lett. B 485 (2000) 208;\\R.
Maartens, Living Rev. Rel. 7 (2004) 7;\\J.D. Bekenstein, Phys. Rev.
D 70 (2004) 083509;\\T. Jacobson, D. Mattingly, Phys. Rev. D 64
(2001) 024028.
\bibitem[5]{5}T.P. Sotiriou, V. Faraoni, \href{http://arxiv.org/abs/0805.1726v2}{arXiv: 0805.1726v2} [gr-qc].
\bibitem[6]{6}S. Nojiri, S.D. Odintsov, \href{http://arxiv.org/abs/hep-th/0601213}{arXiv: 0601213} [hep-th].

\bibitem[7]{7}S. Capozziello, Int. J. Mod. Phys. D 11 (2002) 483.
\bibitem[8]{8}S. Nojiri, S.D. Odintsov, \href{http://arxiv.org/abs/hep-th/0308176}{arXiv: 0308176} [hep-th].
\bibitem[9]{9}S. Nojiri, S.D. Odintsov, \href{http://arxiv.org/abs/hep-th/0307288}{arXiv: 0307288} [hep-th].
\bibitem[10]{10}S. Nojiri, S.D. Odintsov, \href{http://arxiv.org/abs/hep-th/0508049}{arXiv: 0508049} [hep-th].
\bibitem[11]{11}T. Harko, \href{http://arxiv.org/abs/0810.0742}{arXiv: 0810.0742} [gr-qc].
\bibitem[12]{12}O. Bertolami, C.G. Boehmer, T. Harko, F.S.N. Lobo, \href{http://arxiv.org/abs/0704.1733}{arXiv: 0704.1733v2} [gr-qc].
\bibitem[13]{13}J.H. Kung, Phys. Rev. D 52 (1995) 6922;\\J.H. Kung, Phys. Rev.
D 53 (1996) 3017;\\S.E.P. Bergliaffa, Phys. Lett. B 642 (2006)
311;\\ S. Carroll, Spacetime and Geometry: An Introduction to
General Relativity (Addison Wesley, New York) 2004.
\bibitem[14]{14}S.W. Hawking, G.F.R. Ellis, The Large scale structure of space-time (Cambridge University Press, Cambridge 1973).
\bibitem[15]{15}O. Bertolami, M.C. Sequeira, \href{http://arxiv.org/abs/0903.4540}{arXiv: 0903.4540} [gr-qc].
\bibitem[16]{16}J. Santos, J.S. Alcaniz, M.J. Reboucas, F.C. Carvalho, \href{http://arxiv.org/abs/0708.0411}{arXiv: 0708.0411} [astro-ph].
\bibitem[17]{17}M. Visser, Class. Quantum Grav. 21 (2004) 2603, Gen. Relativ.
Gravit. 37 (2005) 1541;\\E.R. Harrison, Nature 260 (1976) 591;\\P.
Landsberg, Nature 263 (1976) 217 .
\bibitem[18]{18}O. Bertolami, T. Harko, F.S.N. Lobo, J. P¨¢ramos, \href{http://arxiv.org/abs/0811.2876}{arXiv: 0811.2876} [gr-qc];\\J.D. Brown, Class. Quant. Grav. 10 (1993)
1579.
\bibitem[19]{19}E. Komatsu, et al., Astrophys. J. Suppl. 180 (2008) 330-376.
\bibitem[20]{20}D. Rapetti, S.W. Allen, M.A. Amin, R.D. Blandford, Mont. Not. R. Soc. 375 (2007) 1510.

\bibitem[21]{21}Alvaro Nunez, Slava Solganik, \href{http://arxiv.org/abs/hep-th/0403159}{arXiv: 0403159} [hep-th].

\bibitem[22]{22}A.D. Dolgov, M. Kawasaki, Phys. Lett. B 573 (2003) 1.

\bibitem[23]{23}B. Bertotti, L. Iess, P. Tortora, Nature 425 (2003) 374.

\bibitem[24]{24} Valerio Faraoni, Phys. Rev. D 76 (2007) 127501.

\bibitem[25]{25} Valerio Faraoni, Phys. Rev. D 74 (2006) 104017.








\end{thebibliography}
\end{document}